# Mapping Researchers with PEOPLEMAP

Jon Saad-Falcon*, Omar Shaikh*, Zijie J. Wang*, Austin P. Wright*, Sasha Richardson†, Duen Horng (Polo) Chau*

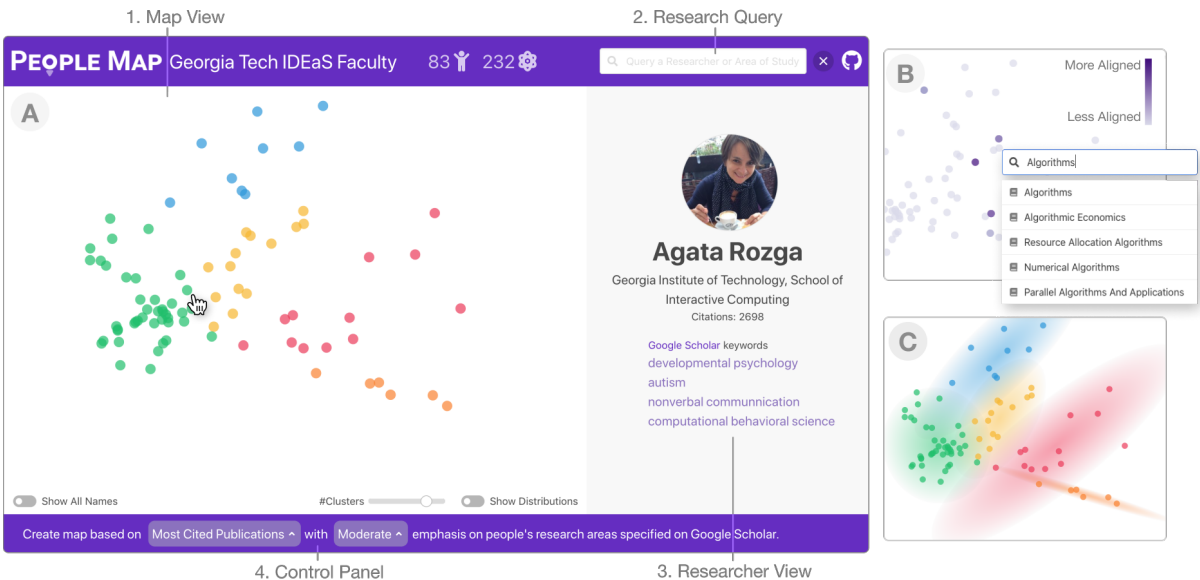

Figure 1: **A.** PEOPLEMAP for 83 Georgia Tech researchers affiliated with the Institute of Data Engineering and Science (IDEaS) based on their research interests and publications. 1. **Map View** visualizes embeddings of researchers generated from their Google Scholar keywords and publication text; each dot represents one researcher. 2. **Research Query** allows searching for researchers and areas of research. 3. **Researcher View** shows detailed information (e.g., affiliation, citation count) of researcher selected in Map View. 4. **Control Panel** for adjusting visualization settings (e.g., show researcher names). **B.** Example search results when querying for the "algorithms" research topic; darker color indicates stronger research alignment between a researcher (dot) and the query topic. **C.** Example researcher clustering results produced by a Gaussian mixture model; cluster distributions shown as ellipses.

## ABSTRACT

Discovering research expertise at universities can be a difficult task. Directories routinely become outdated, and few help in visually summarizing researchers' work or supporting the exploration of shared interests among researchers. This results in lost opportunities for both internal and external entities to discover new connections, nurture research collaboration, and explore the diversity of research.

To address this problem, at Georgia Tech, we have been developing PEOPLEMAP, an open-source interactive web-based tool that uses natural language processing (NLP) to create visual maps for researchers based on their research interests and publications. Requiring only the researchers' Google Scholar profiles as input, PEOPLEMAP generates and visualizes embeddings for the researchers, significantly reducing the need for manual curation of publication information. To encourage and facilitate easy adoption and extension of PEOPLEMAP, we have open-sourced it under the permissive MIT license at https://github.com/poloclub/people-map. PEOPLEMAP has received positive feedback and enthusiasm for expanding its adoption across Georgia Tech.

---

*Georgia Institute of Technology.
 {jonsaadfalcon, oshaikh, jayw, apwright, polo}@gatech.edu
†Fayetteville State University. srichardson@broncos.uncfsu.edu

## 1 INTRODUCTION

Discovering research expertise and potential collaborators at universities can be a difficult task. While manually curated university directories currently fill this role, they are primarily designed for cataloging individuals' affiliation and contact information. Few help in visually summarizing researchers' work or supporting the exploration of shared interests among researchers. Furthermore, such directories routinely become outdated and sometimes provide inaccurate or incomplete information about the researchers as research interests and publication records evolve over time. This results in lost opportunities for both internal and external entities to nurture research collaboration and explore the diversity of research. To address this common issue shared among research institutions, our ongoing work makes the following contributions:

1. **PEOPLEMAP** (Fig. 1A), an open-source interactive web-based tool that employs embeddings generated using natural language processing (NLP) techniques to visually "map out" researchers using their research interests and publications found on the researchers' Google Scholar profiles. Requiring only Google Scholar profiles as input (e.g., their URLs), PEOPLEMAP significantly reduces the need for manual curation of publication information. While existing tools and research have primarily focused on tackling tasks such as recommending research papers and venues to publish at [1, 3, 4], we are working to contribute PEOPLEMAP as one of the first practical tools that helps summarize and visualize researcher interests and expertise.

   To encourage and facilitate easy adoption and extension of PEOPLEMAP, we have open-sourced it under the permissive

MIT licence. PEOPLEMAP's code repository and detailed documentation is available at https://github.com/poloclub/people-map. All generated PEOPLEMAPs are static web applications that can be hosted as standard web pages (e.g., as GitHub pages) without the need for any backend computation servers.

2. **Deployment of PEOPLEMAP: Early Usage and Feedback** To demonstrate the feasibility and generalizability of PEOPLEMAP, we have successfully deployed PEOPLEMAPs for three research units at Georgia Tech: (1) the *Institute of Data Engineering and Science* (IDEaS)[1], with 83 affiliated faculty members, that serves as a unified point to connect researchers with government and industry to advance foundational data science research; (2) the Center of Machine Learning[2] with over 40 core faculty members; and (3) the Department of Chemistry and Biochemistry[3] with 32 faculty members. We have enjoyed positive feedback from leadership teams for this early deployment. Some faculty members are particularly excited about PEOPLEMAP's interactive exploration support and its potential in helping them find colleagues to collaborate with on research projects and grant proposals. Discussion has begun on expanding PEOPLEMAP's adoption across more research units across Georgia Tech.

## 2 PEOPLEMAP SYSTEM DESIGN

PEOPLEMAP's user interface holds four major components:

**Mapping Out Researcher Interests.** The *Map View* component (Fig. 1A-1) visualizes the researcher embeddings, allowing the user to explore the similarities and differences between researchers. These researcher embeddings were generated by first gathering the research interests and publications from each researcher's Google Scholar profile, using the scholarly Python library[4]; this process only requires the researchers' Google Scholar profile URLs. We then concatenate the titles and abstracts of each researcher's publications together; for some configurations of PEOPLEMAP's settings, researcher keywords are also added into their combined document. To normalize these combined documents, non-English characters and stopwords are removed, words are stemmed, and characters are turned lowercase. The collected data is then processed using *term frequency–inverse document frequency* (TFIDF) [2], which allows us to penalize common terms (low TFIDF score) shared by the whole dataset and focus on finding "characteristic" terms that differentiate (high score). The TFIDF-weighting placed on a researcher's Google Scholar keywords can be adjusted using the Control Panel (Fig. 1A-4). Each researcher's embedding becomes a column in a TFIDF matrix, where each row is a term, and the cell value is the term's TFIDF score in the embedding. As there are thousands of terms, we perform principal component analysis (PCA) to reduce the dimensionality. We then perform Gaussian mixture modeling to split the overall distribution of researcher vectors into several different Gaussian distributions; each researcher vector in the Map View (Fig. 1C) are colored according to the distribution they are assigned. These Gaussian distributions are intended to aid users in their analysis of the different fields of study among the researchers.

**Finding Specific Researchers and Areas of Study.** When companies and national labs seek to collaborate with a research institution, they often need to first discover whether any researchers' interests align with theirs, and whether there is a critical mass of researchers that could sustain the research engagement. PEOPLEMAP's *Research Query* tool (Fig. 1A-2) aims to support such discovery. It allows users to search for researchers based on how well their research interests align with the query topic (see Fig. 1B for example search results when querying for the "algorithms" research topic). When a user types in a Google Scholar keyword, PEOPLEMAP visualizes how each researcher aligns with the given topic. To determine this alignment between each researcher and the field of interest, we compute the cosine similarity by using the same TFIDF researcher embeddings described earlier. Fig. 1B shows an example query result, where darker colors indicate stronger research alignment, highlighting those who tend to use the query term proportionally more in their publications than other researchers. This feature helps users more easily assess the scope of research relevance.

**Learning More About a Researcher.** The Researcher View (Fig. 1A-3) shows information related to a researcher's profile when they are highlighted or hovered over in the Map View. This profile information includes: their name, affiliation, position, citation count, Google Scholar profile link, and Google Scholar keywords.

**Calibrating Exploration.** The Control Panel (Fig. 1A-4) allows the user to control various configurations about the Map View component. These tools include: the *Show Distributions* toggle allows the user to display the Gaussian distributions generated in the Map View (Fig. 1C); the *#Clusters* slider allows the user to change the number of distributions generated in the Gaussian mixture model; the *Keywords Emphasis* drop-down allows the user to change the weight placed on each researcher's Google Scholar keywords when creating their TFIDF embedding; and the *Publication Set* dropdown allows the user to change the publications (most highly cited, or most recent) used to generate the researcher embedding.

## 3 CONCLUSION AND ONGOING WORK

As PEOPLEMAP continues to gain adoption, we plan to enhance it by exploring more embedding techniques such as Transformer [5] models like BERT to improve information extraction from researcher datasets, and the visualization of researcher embeddings. Additionally, we will investigate the usability and accuracy of other topic modeling techniques, such as employing non-negative matrix factorization (NMF) [1] to identify research fields of interest in a dataset. This can potentially allow us to enhance the explorability of PEOPLEMAP by providing visualized labels for user-selected clusters.

Additionally, we plan to conduct lab studies to evaluate PEOPLEMAP's usability, and work with administrators and industry partners to better understand how PEOPLEMAP could support their discovery of relevant researchers for their diverse array of research projects. As we test PEOPLEMAP with different research entities, we will better understand how the visualization and embedding techniques may work in different conditions to identify potential constraints, such as dataset size and visualizing research interests.

We look forward to more institutions adopting PEOPLEMAP to complement their directories, so that both internal and external entities can better explore the diversity of their research expertise.

---

[1] https://poloclub.github.io/people-map/ideas/
[2] https://poloclub.github.io/people-map/ml/
[3] https://poloclub.github.io/private-people-map/
[4] https://pypi.org/project/scholarly/